\documentclass[aps,twocolumn,pra,superscriptaddress,reprint]{revtex4-2}
\usepackage[utf8]{inputenc}
\usepackage[T1]{fontenc}

\usepackage{amsmath}
\usepackage{amsfonts}
\usepackage{amssymb}
\usepackage{dsfont}

\usepackage[english]{babel}

\usepackage{physics}

\usepackage{hyphenat}
\usepackage{hyperref}
\usepackage{xcolor}
\usepackage{xspace}
\usepackage{graphicx}

\newcommand{\duc}{d_{\text{uc}}}

\begin{document}

\title{Quantum criticality of the ferromagnetic Dicke-Ising model}
\author{Jan Alexander Koziol}
\email{jan.alexander.koziol@univie.ac.at}
\affiliation{Faculty of Physics, University of Vienna, Boltzmanngasse 5, AT-1090 Vienna, Austria}

\begin{abstract}
    We describe the quantum phase transitions in the ferromagnetic Dicke-Ising model using a Landau theory approach.
    The theory quantitatively captures the change from a second- to a first-order transition between the normal and superradiant phases through a tricritical point.
    We identify virtual nearest-neighbor double spin-flip processes as the crucial mechanism responsible for this behavior.
    The tricritical point constitutes a quantum phase transition above the upper critical dimension.
    We discuss the modifications to finite-size scaling required for the correct interpretation of numerical data at the tricritical point.
	Our results emphasize the need for adapted finite-size scaling forms in all-to-all interacting quantum systems and establish the ferromagnetic Dicke-Ising model as a paradigmatic platform for quantum phase transitions above the upper critical dimension, encompassing both standard $\phi^4$ criticality and beyond.
\end{abstract}

\maketitle

\section{Introduction}

Analog quantum simulation platforms form a bridge between condensed matter physics, statistical mechanics, and quantum optics \cite{Browaeys2020,Monroe2021,Chomaz2022,Defenu2023}.
On the one hand, highly controllable atomic and molecular quantum simulators can realize paradigmatic models of quantum many-body physics as well as realistic models of materials \cite{Aidelsburger2013,Bernien2017,Lienhardt2018,Scholl2021,Chen2023,Su2023,Michel2024,Leclerc2026}.
On the other hand, spin models inspired by quantum optics are introduced and investigated using established methods of condensed matter physics \cite{Hepp1973A,Hepp1973B,Leggett1987,Dimer2007,Baumann2010,Haas2014,Klinder2015,Landig2016,Zhiqiang2017,Zhang2018,Defenu2023}.

A prime example of the latter is the Dicke-Ising model \cite{Lee2004,Gammelmark2012,Zhang2013,Zhang2014,Gelhausen2016,Schuler2020,Rohn2020,Schellenberger2024,Lenk2022,Puel2024,Roche2024A,Roche2024B,Koziol2025}.
This model extends the Dicke model \cite{Dicke1954,Hepp1973A,Hepp1973B,Wang1973}, which describes the coupling of $N$ two-level systems (spin-$1/2$) to a single bosonic mode, by introducing a lattice structure and additional Ising interactions.
The Dicke-Ising model is a paradigmatic model for investigating the competition between long-range boson-mediated interactions and short-range spin-spin interactions \cite{Lee2004,Zhang2013,Zhang2014,Gelhausen2016,Schuler2020,Rohn2020,Schellenberger2024,Puel2024,Roche2024A,Roche2024B,Koziol2025}.
The all-to-all nature of the spin-boson coupling often leads to a strongly simplified dynamical behavior in open and closed Dicke systems \cite{Hepp1973A,Hepp1973B,Wang1973,Vidal2006,Gelhausen2016,Larson2017,Schellenberger2024,Koziol2025,Langheld2024,Bassler2025,Rosario2025,Zhang2025}.
The inclusion of additional short-range interactions is a way to introduce non-trivial behavior into the system \cite{Zhang2014,Gelhausen2016,Schellenberger2024,Langheld2024,Koziol2025,Mendonca2025,Prazeres2026,Winter2026}.
Competing Ising interactions give rise to structured normal and superradiant phases \cite{Zhang2014,Gelhausen2016,Langheld2024,Koziol2025,Leibig2026}.
Nevertheless, normal phases of Dicke-Ising models are exact product states in the thermodynamic limit, and their low-energy properties can be understood in terms of standard multi-magnon-mode Dicke models without Ising interactions \cite{Schellenberger2024,Koziol2025}.
This implies that the generic transition between a normal and a superradiant phase is the same second-order phase transition as in the Dicke model when both phases share the same spatial symmetries, and a first-order phase transition otherwise \cite{Koziol2025}.
The notable exception to this generic rule is the quantum phase transition in the ferromagnetic Dicke-Ising model studied in this work.
Here, the phase transition changes from second order to first order between phases with the same spatial symmetry \cite{Rohn2020,Langheld2024,Leibig2026}.
In this work, we will provide a quantitative theoretical explanation of this phenomenon and identify it as a genuine effect of the short-range Ising interactions in the model.

Finite-size scaling above the upper critical dimension (UCD, $\duc$) is a subtle aspect of continuous phase transitions 
\cite{Binder1985,Binder1985B,Luijten1997,Kenna2013,FloresSola2015,FloresSola2016,Koziol2021,Berche2022,Langheld2022,Adelhardt2024} that becomes particularly important for generalized Dicke models \cite{Langheld2024}.
Above their UCD, phase transitions exhibit mean-field critical exponents, but their finite-size scaling is not described by the standard scaling forms \cite{Berche2022,Langheld2022}.
Instead, dangerous irrelevant variables \cite{Fisher1983} modify the scaling relations and the finite-size scaling forms used to extract critical exponents from numerical data for finite systems \cite{Berche2022,Langheld2022}.
For conventional short-range quantum systems, this regime is difficult to access.
For instance, the UCD of the canonical $O(n)$ quantum universality classes, including Ising, XY, and $O(3)$ transitions, is three \cite{Suzuki1976,SachdevBook}.
Observing finite-size scaling above the UCD would therefore require at least four spatial dimensions.
Long-range quantum lattice models motivated by quantum simulators have changed this perspective, since algebraically decaying interactions can effectively lower the UCD and make modified finite-size scaling relevant in low-dimensional systems \cite{Koziol2021,Langheld2022,Zhao2023,Adelhardt2023,Song2024,Adelhardt2025,Adelhardt2026}.
Dicke models provide an even more direct route.
As all-to-all interacting quantum systems, their continuous ground-state phase transitions naturally lie in the regime above the UCD \cite{Langheld2024,Koziol2025}.
Finite-size scaling above the UCD is therefore not only a theoretical issue but will also become experimentally relevant in quantum simulation platforms \cite{Dimer2007,Zhiqiang2017,Zhang2018}.

In this work, we provide a comprehensive theory of the phase transitions in the ferromagnetic Dicke-Ising model. 
In Sec.~\ref{sec:FerromagneticDickeIsingModel}, we introduce the model.
We summarize the existing literature, focusing on the quantum phase diagrams and phase transitions.
In Sec.~\ref{sec:TricriticalMeanFieldTheory}, we discuss how Landau theory describes the change from second- to first-order transitions through a tricritical point.
In Sec.~\ref{sec:derivation}, we explicitly derive the Landau free energy for the Dicke-Ising model.
Furthermore, we derive quantitative predictions for finite-size scaling at the tricritical point in Sec.~\ref{sec:FiniteSizeScalingA}.
In Sec.~\ref{sec:FiniteSizeScaling}, we use quantum Monte Carlo data from the publicly available repository \cite{Langheld2024,Langheld2025Data} to confirm the tricritical mean-field point through an explicit finite-size scaling analysis.
We conclude in Sec.~\ref{sec:fin} by discussing our results in the context of (i) short-range interactions modifying the universal behavior of all-to-all interacting Dicke models and (ii) the tricritical mean-field point as an example of finite-size scaling above the UCD beyond standard $\phi^4$ theory.

\section{Ferromagnetic Dicke-Ising model and relations to previous work}
\label{sec:FerromagneticDickeIsingModel}

The Dicke-Ising model extends the paradigmatic Dicke model \cite{Dicke1954,Hepp1973A,Hepp1973B,Wang1973} by including Ising interactions \cite{Lee2004,Gammelmark2011,Zhang2013,Zhang2014,Gelhausen2016,Schuler2020,Rohn2020,Schellenberger2024,Puel2024,Roche2024A,Roche2024B,Koziol2025}.
In the Dicke model, $N$ two-level systems with level splitting $2\epsilon$ are coupled to a single bosonic mode with energy $\omega$.
The coupling strength is $g/\sqrt{N}$.
In contrast to the Dicke model, the spatial arrangement of the two-level degrees of freedom plays an important role in the Dicke-Ising model.
We arrange the degrees of freedom on lattices.
The Dicke-Ising Hamiltonian with ferromagnetic nearest-neighbor Ising interactions $(J>0)$ is
\begin{equation}
	\begin{split}
			H=&+\omega a^\dagger a + \frac{g}{\sqrt{N}}(a^\dagger+a)\sum_{i}\sigma_i^x\\
			  &-\epsilon\sum_{i}\sigma_i^z-J\sum_{\langle i,j\rangle} \sigma_i^z\sigma_j^z \ .
\end{split}
\label{eq:DickeIsingModel}
\end{equation}
The bosonic creation and annihilation operators are $a^\dagger$ and $a$.
The two-level degrees of freedom are described by the Pauli operators $\sigma_i^{(x/z)}$.
Without loss of generality, we assume $\epsilon>0$ \footnote{Solutions for $\epsilon<0$ are obtained by $\sigma^z_i\rightarrow -\sigma_i^z$.}.
The Hamiltonian has a $\mathbb{Z}_2$ symmetry: the transformation $(a,\sigma_i^x)\rightarrow (-a,-\sigma_i^x)$ leaves the Hamiltonian invariant.

\begin{figure*}[t]
	\centering
	\includegraphics[width=\textwidth]{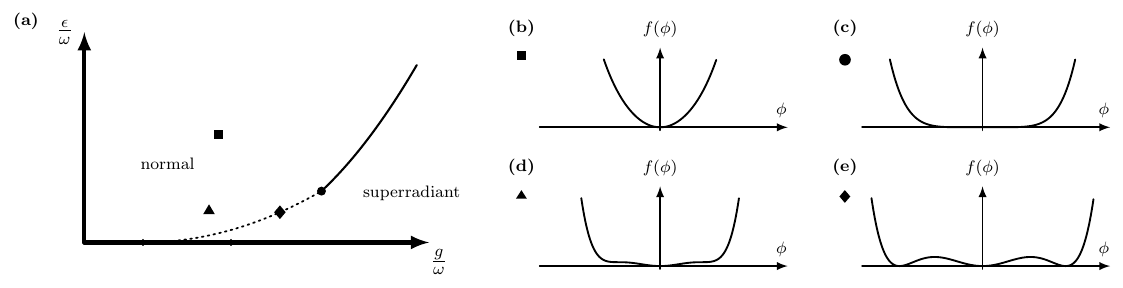}
	\caption{
\textbf{(a)} Schematic phase diagram of the ferromagnetic nearest-neighbor Dicke-Ising model at fixed $J/\omega$.
The solid line marks the second-order Dicke transition line.
The tricritical point is marked by the black dot.
The dotted line marks the first-order transition.
\textbf{(b)}-\textbf{(e)} Landau free-energy density of the $\phi^6$ theory in Eq.~\eqref{eq:phisix}: 
\textbf{(b)} Symmetric phase with $r,u,v>0$;
\textbf{(c)} Tricritical point with $r=u=0$ and $v>0$;
\textbf{(d)} Symmetric phase with $r,v>0$ and $u<0$; 
\textbf{(e)} First-order transition with $r,v>0$ and $u<0$.
Markers connect the free-energy profiles to corresponding points in panel \textbf{(a)}.
}
	\label{fig:TentativePhaseDiagram}
\end{figure*}

In the limit $g=0$, the ground state is a $z$-polarized product state \cite{Zhang2013,Rohn2020,Langheld2024,Leibig2026}.
In the thermodynamic limit, this state remains an exact ground state for finite $g$ up to the phase transition at $g_c$ \cite{Schellenberger2024}. 
Beyond the transition, the system enters a superradiant phase characterized by a finite boson density in the ground state \cite{Zhang2013,Rohn2020,Langheld2024,Leibig2026}.
We refer to the phase at $g<g_c$ as the normal phase and the phase at $g>g_c$ as the superradiant phase \cite{Zhang2013,Rohn2020,Langheld2024,Leibig2026}.
The superradiant phase breaks the $\mathbb{Z}_2$ symmetry of the Hamiltonian \cite{Zhang2013,Rohn2020,Langheld2024,Leibig2026}.
For $\epsilon>J$, a second-order phase transition is reported \cite{Zhang2014,Schellenberger2024,Langheld2024,Leibig2026,Mendonca2025} (see Fig.~\ref{fig:TentativePhaseDiagram}).
However, for $\epsilon < J$, several studies indicate a first-order phase transition between these two states \cite{Rohn2020,Langheld2024,Leibig2026,Sur2026} (see Fig.~\ref{fig:TentativePhaseDiagram}).

The key question is the origin of this change in critical behavior.
Product-state mean-field theory \cite{Zhang2014} and the low-energy theory of polariton condensation \cite{Schellenberger2024} fail to capture this change in critical behavior.
In contrast, the Landau theory derived in this work describes this phenomenon quantitatively and identifies it as a genuine effect of the Ising interactions.

Several related works discuss first- and second-order quantum phase transitions in the transverse-field Dicke-Ising chain \cite{Gammelmark2011,Guo2025,Sur2026,Otake2026}.
One difference between this work and the related studies lies in the relative orientation of the Ising interaction, the light-matter coupling, and the level splitting.
Another difference is the general applicability of the perturbative calculations outlined in this work to non-integrable models.
Our findings at $\epsilon=0$ for a chain geometry connect to these studies.
Also, the finite-size scaling analysis discussed here is relevant for the models discussed in \cite{Gammelmark2011,Guo2025,Sur2026}.

We note that one numerical study claims that the quantum phase transition in the Dicke-Ising model remains a second-order Dicke transition throughout the phase diagram \cite{Mendonca2025}.
In this study, the authors use a variational unitary transformation accompanied by density matrix renormalization group (DMRG) calculations \cite{Mendonca2025}.
As pointed out in the comment \cite{Hoermann2025}, the work \cite{Mendonca2025} contradicts other quantitative studies of the phase diagram and criticality of the model \cite{Gammelmark2011,Guo2025,Rohn2020,Langheld2024,Leibig2026}.

\section{Landau theory with a tricritical point}
\label{sec:TricriticalMeanFieldTheory}
In this section we follow Ref.~\cite{Chaikin1995}.
It is well established that first-order transitions can occur even when symmetry prohibits odd-order terms in the Landau theory.
Let us consider the Landau free-energy density $f$ as a function of the order-parameter density $\phi$
\begin{equation}
	\label{eq:phisix}
	f(\phi)= r \phi^2 + u\phi^4 + v\phi^6 
\end{equation}
with $v>0$ for stability.
For the standard second-order Dicke transition, $u>0$, so the $v$ term can be neglected.
If $r \propto g_c-g > 0$, the free energy is minimized at $\phi = 0$, and the system is in the symmetric (normal) phase. 
For $r \propto g_c-g < 0$, the system is in the superradiant phase.
Now if $u<0$, the sixth-order term is required to maintain stability.
If $u<0$, secondary minima appear symmetrically around $\phi=0$ (see Fig.~\ref{fig:TentativePhaseDiagram}).
When these secondary minima cross the line $f(\phi)=0$ at $r>0$, the system undergoes a first-order transition (see Fig.~\ref{fig:TentativePhaseDiagram}).
If $r=u=0$ simultaneously, the transition remains second order, but the theory is stabilized by the $\phi^6$ term (see Fig.~\ref{fig:TentativePhaseDiagram}).
As a consequence, the critical exponents and the UCD differ at this so-called tricritical point since there is no $\phi^4$ term present.
The static critical exponents are $\alpha=1/2$, $\beta = 1/4$, $\gamma = 1$, and $\delta = 5$ \cite{Chaikin1995}.
Power-counting arguments show that the $\duc$ of a $\phi^6$-stabilized theory is lower than that of a $\phi^4$-stabilized theory \cite{SachdevBook}.

\section{Derivation of the effective Landau theory}
\label{sec:derivation}
In this section, we derive a Landau theory for the quantum-critical behavior of the Dicke-Ising model.
We begin by following an argument that maps the ground-state behavior of the Dicke-Ising model onto an all-to-all interacting pure spin model.
Starting with the Dicke-Ising model in Eq.~\eqref{eq:DickeIsingModel}, we perform a displacement transformation \cite{Langheld2024}
\begin{equation}
		U=\exp\left(\frac{g}{\sqrt{N}\omega} (a^\dagger-a)X\right)\quad\text{with}\quad X=\sum_i \sigma_i^x \ .
\end{equation}
The transformed Hamiltonian reads
\begin{equation}
		\tilde H = U H U^\dagger = \omega a^\dagger a - \frac{g^2}{\omega N} X^2 + U H_{0} U^\dagger 
\end{equation}
with
\begin{equation}
	H_0 = -\epsilon \sum_i\sigma_i^z -J\sum_{\langle i,j \rangle}\sigma_i^z \sigma_j^z \ .
\end{equation}
It was shown in Ref.~\cite{Langheld2024} that
\begin{equation}
		U H_{0} U^\dagger = -\epsilon \sum_{i}\sigma_i^z - J \sum_{\langle i,j\rangle}\sigma_i^z \sigma_j^z + \mathcal{R}
\end{equation}
where the remainder $\mathcal{R}$ scales subextensively.
Therefore, $\mathcal{R}/N$ goes to zero in the thermodynamic limit when considering the ground-state density.

Understanding the ground-state behavior of the Dicke-Ising model reduces to studying the ground state of the following single-axis Lipkin-Meshkov-Glick model with additional nearest-neighbor Ising interactions
\begin{align}
\label{eq:pub_H} H&=-\epsilon\sum_i \sigma_i^z-J\sum_{\langle i,j\rangle}\sigma_i^z\sigma_j^z+\frac{\lambda}{N}\sum_{i,j}\sigma_i^x\sigma_j^x\\
\label{eq:pub_H0} &=H_0+\frac{\lambda}{N}X^2
\end{align}
where $\lambda=-\frac{g^2}{\omega}$.
At this point, it is instructive to note that the $\mathbb{Z}_2$ symmetry breaking associated with the superradiant phase transition is encoded in magnetic order along the $x$ direction.
Alternative approaches to establish an equivalence between Lipkin-Meshkov-Glick matter models and generalized Dicke models can be found in \cite{Reslen2005,Lenk2022,Roche2024A,Roche2024B}.

The model studied here belongs to the class of models with separable interactions and additional local terms \cite{Tindemans1975}.
For this class of models, an extensive theoretical framework exists that maps the problem onto a self-consistent Hamiltonian \cite{Tindemans1975,denOuden1976A,denOuden1976B,Perk1977}.
This approach is, for example, exploited in the numerical study of Ref.~\cite{Leibig2026}.
Furthermore, the approach of Refs.~\cite{Tindemans1975,denOuden1976A,denOuden1976B,Perk1977} rigorously maps the analysis of the Landau free energy onto a model that is linear in the operator $X$.

To derive the Landau theory of the quantum phase transitions in the ferromagnetic Dicke-Ising model, we analyze the zero-temperature onset of $x$-order in the single-axis Lipkin-Meshkov-Glick model in Eq.~\eqref{eq:pub_H0}.
We characterize the instability of the normal state towards a finite $m_x$ magnetization through the response of $H_0$ to a uniform probe field $h_x$ \cite{Perk1977,Capel1979},
\begin{equation}
\bar H(h_x)=H_0-h_x X\ .
\label{eq:pub_probe}
\end{equation}
The strategy is to derive the Landau free-energy density $\bar{f}(m_x)$ from Eq.~\eqref{eq:pub_probe} \cite{Capel1979}.
We introduce the effect of the all-to-all interaction by adding $\lambda m_x^2$ afterwards \cite{Capel1979}.
The ground-state energy of Eq.~\eqref{eq:pub_probe} is in general not accessible analytically \footnote{The notable exception is the model with $\epsilon=0$ on the chain. Here an analytical solution is available \cite{Pfeuty1970,Pfeuty1971A}}.
Therefore, we will resort to a perturbative expansion in $h_x$.
A straightforward calculation \cite{Takahashi1977} then yields the ground-state energy density of the short-range model on a regular lattice with coordination number $z$
\begin{equation}
\bar e(h_x)= e_0 -\frac{h_x^2}{2A} +\frac{|\epsilon|-J}{8A^3B}\,h_x^4 +O(h_x^6)\ ,
\label{eq:pub_eh}
\end{equation}
where $e_0=-|\epsilon|-zJ/2$, $A=|\epsilon|+2J$, and $B=|\epsilon|+J$. 
The corresponding magnetization is
\begin{equation}
m_x= -\frac{\partial \bar e}{\partial h_x} = \frac{h_x}{A} -\frac{|\epsilon|-J}{2A^3B}\,h_x^3 +O(h_x^5)\ .
\label{eq:pub_mh}
\end{equation}

We perform a Legendre transformation of the short-range energy, $\bar f(m_x)=\bar e(h_x)+h_x m_x$, where $h_x$ is eliminated in favor of $m_x$ by inverting the series in Eq.~\eqref{eq:pub_mh}.
This yields
\begin{equation}
\bar f(m_x)= e_0 +\frac{A}{2}m_x^2 +\frac{A(|\epsilon|-J)}{8B}m_x^4+O(m_x^6)\ .
\label{eq:pub_f0}
\end{equation}
The Landau free energy in Eq.~\eqref{eq:pub_f0} encodes, through its gradient at a given magnetization, the field strength required to obtain that magnetization.
Adding the collective contribution $\lambda m_x^2$ of the original model back into the theory, we obtain the effective Landau free energy of the full model \cite{Perk1977,Capel1979},
\begin{equation}
f_{\mathrm{eff}}(m_x)= e_0 +\left(\lambda+\frac{A}{2}\right)m_x^2 +u\,m_x^4 +O(m_x^6) \ ,
\label{eq:pub_feff}
\end{equation}
with
\begin{equation}
u=\frac{A(|\epsilon|-J)}{8B}=\frac{|\epsilon|}{8}-\frac{zJ^2}{8\bigl(|\epsilon|+(z-1)J\bigr)}\ .
\label{eq:pub_u}
\end{equation}
The term $|\epsilon|/8$ in Eq.~\eqref{eq:pub_u} corresponds to the naive product-state mean-field result. 
The second term is the correction generated by the Ising interactions.
Its origin can be traced back to the fourth-order term in the perturbative ground-state energy expansion Eq.~\eqref{eq:pub_eh}.
It can be attributed to perturbative processes with intermediate states with two neighboring spins flipped.
The interaction therefore leaves the quadratic instability line unchanged from the product-state mean-field result but renormalizes the quartic coefficient in a non-trivial way.
In particular, $u>0$ for $|\epsilon|>J$, so the transition remains a Dicke transition; 
at $|\epsilon|=J$, the fourth-order coefficient vanishes, and the sixth-order term in the Landau theory becomes relevant;
for $|\epsilon|<J$, the fourth-order coefficient becomes negative, indicating that the second-order transition is preempted by a first-order transition.
See Fig.~\ref{fig:TentativePhaseDiagram} for a visualization.

\section{Finite-size scaling predictions}
\label{sec:FiniteSizeScalingA}

Next, we connect the Landau theory to predictions for finite-size scaling.
For the standard non-interacting Dicke model with $N$ two-level systems, the order parameter is commonly analyzed using
\begin{equation}
	\Phi_N(r\propto g_c-g)=N^{-\beta/\nu}\phi(N^{1/\nu}r) \ .
    \label{eq:NaiveFSS}
\end{equation}
This gives $\nu=3/2$ \cite{Lambert2004,Reslen2005,Vidal2006,Liu2009,Langheld2024}.
Taken literally, this is puzzling: the standard Dicke model has no spatial structure, yet it exhibits a finite exponent $\nu$ governing the divergence of the correlation length.

The resolution is that the Dicke transition is understood as a quantum phase transition above its upper critical dimension in an all-to-all interacting Ising model \cite{Langheld2024}.
Consequently, finite-size scaling does not directly measure a conventional correlation-length exponent, but the modified exponent $\nu'=\duc\nu$ \cite{Langheld2022,Berche2022} \footnote{In the literature the usual notation is $\nu'=\duc\nu/d$ and the scaling form Eq.~\eqref{eq:ModifiedFSS} is expressed in terms of the linear system size $L$. Here, we use $L^d=N$ to express the scaling forms in terms of $N$. Both notations are equivalent.}.
For the Dicke transition, $\nu'=3/2$ \cite{Langheld2024}, and the resulting scaling form is \cite{Langheld2022,Langheld2024}
\begin{equation}
	\Phi_N(r\propto g_c-g)=N^{-\beta/\nu'}\phi(N^{1/\nu'}r)\ .
    \label{eq:ModifiedFSS}
\end{equation}
Thus, the exponent extracted from finite-size scaling of Dicke transitions should be interpreted as $\nu'$, rather than a true spatial correlation-length exponent.

The tricritical point of the ferromagnetic Dicke-Ising model follows the same reasoning, but with a different critical theory.
At this point, the quartic Landau coefficient vanishes, and the leading stabilizing term is of sixth order in the order parameter.
The static critical exponents are therefore the mean-field tricritical values, $\alpha=1/2$, $\beta=1/4$, $\gamma=1$, and $\delta=5$ \cite{Chaikin1995}.
A power-counting analysis of the corresponding long-range $\phi^6$ theory, summarized in Appendix~\ref{app:powercounting}, yields the modified finite-size exponent $\nu'=\duc\nu=1$.

In summary, finite-size scaling theory for transitions above the UCD provides the central framework of interpretation of the finite-size numerics in generalized Dicke models.
The ferromagnetic Dicke-Ising model realizes a tricritical quantum phase transition with finite-size scaling governed by a long-range $\phi^6$ theory above its UCD.
It therefore requires an extension of the finite-size scaling framework known from the Dicke transition beyond standard $\phi^4$ criticality. 
This makes the model a natural testbed for the modified scaling at tricritical $\phi^6$ quantum phase transitions above the UCD.

\section{Comparison with numerical data}
\label{sec:FiniteSizeScaling}
In the following, we compare our analytical predictions with the world-line quantum Monte Carlo data of Ref.~\cite{Langheld2024}, which are publicly available in Ref.~\cite{Langheld2025Data}.
We also compare the predicted position of the tricritical point with the thermodynamic-limit estimate of Ref.~\cite{Leibig2026}.

Our theory predicts the tricritical point at $|\epsilon|=J$.
This prediction agrees with the numerical results for the chain and the square lattice \cite{Langheld2024,Leibig2026}.

\begin{figure}[t]
	\centering
	\includegraphics[]{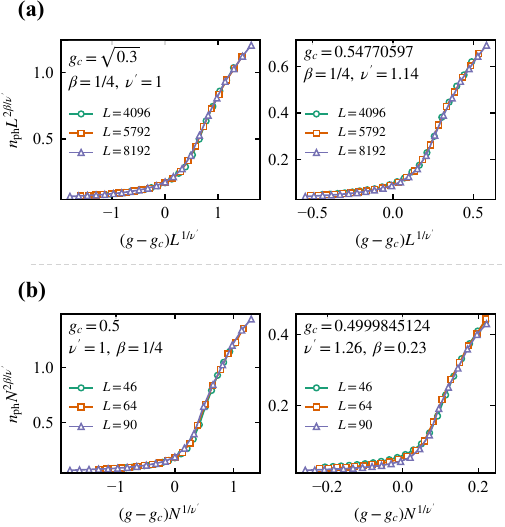}
	\caption{
    Comparison of data collapses of the photon-number-density order parameter at the tricritical point for the chain \textbf{(a)} and square lattice \textbf{(b)}.
    The data are taken from Ref.~\cite{Langheld2025Data}.
    Left panels use the tricritical mean-field values of $g_c$, $\beta$, and $\nu'$.
    Right panels use the freely fitted critical parameters from Ref.~\cite{Langheld2024}.
}
	\label{fig:PlotCollapse}
\end{figure}

Ref.~\cite{Langheld2024} finds the expected Dicke transition at large $|\epsilon|$.
At the tricritical point, Ref.~\cite{Langheld2024} reports $\beta=0.25(1)$ and $\nu'=1.14(3)$ for the chain, and $\beta=0.23(1)$ and $\nu'=1.26(4)$ for the square lattice.
The values of $\beta$ agree with tricritical mean-field theory.
The values of $\nu'$ are larger than our prediction $\nu'=1$.
We attribute this difference to crossover effects in finite-size scaling.
The fits in Ref.~\cite{Langheld2024} used $\beta$, $\nu'$, and $g_c$ as free parameters.
Such fits can smooth sharp changes between scaling regimes, as also observed in long-range transverse-field Ising models \cite{Koziol2021,Langheld2022}.
We therefore reanalyze the raw data of Ref.~\cite{Langheld2025Data}.
We compare two data collapses at the tricritical point.
The first collapse uses the tricritical mean-field values of $g_c$, $\beta$, and $\nu'$.
The second collapse uses the fitted exponents of Ref.~\cite{Langheld2024}.
Fig.~\ref{fig:PlotCollapse} shows the result.
Both collapses have similar visual quality.
A pairwise-deviation score gives a preference for the tricritical parameters.
Thus, the available data are consistent with $\nu'=1$.
Details on the pairwise-deviation analysis are presented in Appendix~\ref{app:DeviationAnalysis}.

We next compare the predicted first-order transition line with the numerical data.
The Landau free energy gives an approximate estimate because higher-order terms shift the onset of the side minima.
For the ferromagnetic chain, we derive the Landau free energy up to eighth order in Appendix~\ref{app:landau}.
Fig.~\ref{fig:plot3} compares the onset of side minima with the Monte Carlo transition points.
The Landau theory reproduces the qualitative shape of the first-order line.
The agreement improves when higher-order terms are included.
The deviation grows away from the tricritical point.

The point $\epsilon=0$ provides an additional check.
At this point, the construction of the free energy can be carried out exactly using the analytical solution of the transverse-field Ising chain
\cite{Pfeuty1970,Pfeuty1971A}.
This gives an onset of the side minima at $g_c/\omega=0.4090436682$ for the parameters of Fig.~\ref{fig:plot3}.
This value agrees perfectly with \cite{Langheld2024,Langheld2025Data}.

All these comparisons fully support the derived theory.
The position of the tricritical point agrees perfectly with numerical data.
The order-parameter exponent $\beta$ at this point agrees with tricritical scaling.
The analysis of the data collapses is consistent with $\nu'=1$.
Further, the first-order line follows the trend predicted by the Landau theory and can be estimated with increasing precision by including higher-order terms.

\begin{figure}
	\centering
	\includegraphics[width=\columnwidth]{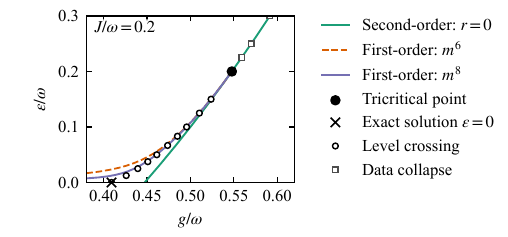}
	\caption{
    Comparison of numerical phase-transition points for the ferromagnetic nearest-neighbor Dicke-Ising chain \cite{Langheld2024,Langheld2025Data} with the Landau theory predictions.
    The exact point at $\epsilon=0$ follows from the transverse-field Ising chain.
    The condition $r=0$ gives the continuous transition for $\epsilon \geq J$.
	Estimates for the first-order transition points from Landau theory including sixth- and eighth-order terms are presented.
}
	\label{fig:plot3}
\end{figure}

\section{Discussion and summary}
\label{sec:fin}
The Dicke-Ising model exhibits a quantum phase transition between a ferromagnetic normal phase and a uniform superradiant phase. 
Depending on the system parameters, the transition is either second-order or first-order, with a tricritical point separating the two regimes \cite{Rohn2020,Langheld2024,Leibig2026}.
A Landau theory that includes a tricritical point explains this scenario quantitatively.
The central idea is that a continuous phase transition changes into a first-order transition when the coefficient of the quartic term in the Landau free energy changes sign.
We explicitly derived the Landau free energy for the ferromagnetic Dicke-Ising model and predicted the position of the tricritical point and the associated critical exponents.
We could attribute the sign change in the Landau free energy explicitly to a contribution of the short-range Ising interaction in our perturbative approach.

A careful interpretation of extracted critical exponents is crucial when performing finite-size scaling analyses of numerical data.
We outlined how the Dicke transition and the tricritical point can be understood as quantum phase transitions above the UCD and how the resulting critical exponents should be interpreted.
A comparison with existing numerical results \cite{Langheld2024,Langheld2025Data,Leibig2026} supports the predictions.

The critical theory of the Dicke-Ising model provides an example of how short-range interactions can alter the quantum critical behavior of all-to-all interacting systems.
Previous studies have shown that polariton condensation in the normal phases of short- and long-range Dicke-Ising models always leads to Dicke-type phase transitions \cite{Schellenberger2024,Koziol2025}.
The mechanism outlined in this work shows that higher-order virtual processes in our perturbative expansion modify the quartic term in the Landau free energy, thereby driving the tricritical point and the first-order transition.

The perturbative approach to obtain a Landau free energy presented in this work relies on a low-field expansion of matter models against a transverse field.
It can be applied to any suitable gapped matter models on arbitrary lattices.
This is especially appealing since series expansions are a well-established tool in the study of spin models and high-order series exist for many relevant systems \cite{Oitmaa2006}.

The applicability of the Landau theory with a tricritical point makes the ferromagnetic Dicke-Ising model a testbed for finite-size scaling above the UCD, both for standard $\phi^4$ theory and beyond.
Deriving the critical exponents at the tricritical point requires consideration of a $\phi^6$ theory.

This suggests that future experimental realizations of quantum phase transitions in Dicke models, especially in the ferromagnetic Dicke-Ising model, could provide a platform for probing finite-size scaling above the UCD \cite{Dimer2007,Gelhausen2016,Zhiqiang2017,Zhang2018}.
Finite-size scaling data obtained for different system sizes could bridge the gap between abstract concepts in the theory of quantum phase transitions \cite{Berche2022,Langheld2022}, such as dangerous irrelevant variables \cite{Fisher1983,SachdevBook,Berche2022} required for deriving the correct scaling forms, and state-of-the-art quantum simulation experiments \cite{Dimer2007,Gelhausen2016,Zhiqiang2017,Zhang2018}.
The finite-size scaling analysis outlined in this work is also applicable to other reported tricritical points in related Dicke-Ising models \cite{Gammelmark2011,Sur2026}, as well as to more general models with separable interactions that exhibit mean-field phase transitions \cite{Tindemans1975,denOuden1976A,denOuden1976B,Perk1977,Capel1979}.

\section{Acknowledgments}
This research was funded in whole or in part by the Austrian Science Fund (FWF) [10.55776/COE1, 10.55776/F101200] and the European Union (NextGenerationEU).
I thank Andreas Nunnenkamp for his guidance on the structure and content of the manuscript, as well as for fruitful discussions.
I also thank Anja Langheld for fruitful ongoing discussions on the topic, for help with the data \cite{Langheld2024,Langheld2025Data}, and for many helpful comments on the manuscript.

%

\newpage
\onecolumngrid

\appendix

\section{Power counting of the long-range $\phi^6$ theory}
\label{app:powercounting}
In this Appendix, we briefly introduce the power counting arguments to derive $\nu'=1$ at the tricritical point.
We consider the effective partition function to describe the quantum phase transition in ferromagnetic long-range transverse-field Ising models with a decay exponent $d+\sigma$ (where $d$ is the spatial dimension of the system) \cite{SachdevBook,Dutta2001,Adelhardt2024}
\begin{align}
		\mathcal{Z}&=\int \mathcal{D}\phi(x,\tau) \ e^{-\mathcal{S}_\phi} \\
		\label{eq:sraction}
		\mathcal{S}_\phi&=\int \mathrm d^dx\int^\beta_0 \mathrm d\tau \ [\{g(\partial_\tau\phi)^2+(\nabla_x^{\sigma/2}\phi)^2+r\phi^2\}+u\phi^4+v\phi^6] \ .
\end{align}
At criticality, the system is scale-free; therefore, for distances $x\gg a$ that are much larger than the lattice spacing, the theory should be invariant under a scaling transformation of coordinates in space and time
\begin{align}
		x \ \rightarrow \ &x^\prime=x/b \\
		\tau \ \rightarrow \ &\tau^\prime=x/b^z
\end{align}
with $b$ being a rescaling factor \cite{SachdevBook}.
The scaling dimensions to keep the action scale invariant are obtained by power counting \cite{SachdevBook},
\begin{align}
		[\mathrm{d}^dx] &= -d \\
		[\mathrm{d}\tau] &= -z = -\sigma/2 \\
		[\nabla_x^{\sigma/2}] &= \sigma/2 \\
		[\partial\tau] &= z = \sigma/2\\
		[\phi] &= (d-z)/2 = (d-\sigma/2)/2\\
		[r] &= \sigma \\
		[u] &=  3\sigma/2-d\\
		[v] &=  2\sigma-2d\ .
\end{align}
For a mean-field transition, the scaling dimension $[r]=1/\nu$ is directly associated with the correlation length exponent \cite{SachdevBook,Dutta2001}.
In the case of standard $\phi^4$ theory, the dimension at which $[u]=0$ is used to define the UCD \cite{SachdevBook,Dutta2001}.
At the tricritical point, where $u=0$, the UCD is defined by the dimension at which the scaling dimension of the next-order stabilizing term becomes zero \cite{SachdevBook}.
Therefore, $\duc=\sigma$; this leads to $\nu'=\duc \nu=1$ for quantum phase transitions above the UCD.

\section{Deviation analysis of data collapses}
\label{app:DeviationAnalysis}
In this Appendix, we introduce the definition of the pairwise deviation score and provide the numerical values for the considered cases.
To obtain this score, we restrict all collapsed curves to their common overlap interval in $x$, interpolate each curve onto the $x$-values of the others, and compute the root-mean-square vertical deviation between all pairs of curves. 
Averaging these pairwise deviations and normalizing by the total $y$-range in the common window yields a dimensionless collapse score $Q$, where smaller values correspond to a better data collapse.
For the chain, we obtain $Q=0.007767$ for the tricritical collapse and $Q=0.008833$ for the fitted collapse.
For the square lattice, we obtain $Q=0.013253$ for the tricritical collapse and $Q=0.017409$ for the fitted collapse.
This indicates that the data collapse using the fixed tricritical-point critical values is actually better than the collapse with the fitted values.

\section{Eighth-order Landau expansion for the ferromagnetic chain}
\label{app:landau}
In this Appendix, we summarize the zero-temperature Landau expansion of the ferromagnetic chain.
We define
\begin{equation}
A=|\epsilon|+2J\qquad B=|\epsilon|+J\qquad C=3|\epsilon|+2J\qquad D=2|\epsilon|+J\ .
\end{equation}
\noindent The perturbative expansion of ground-state energy density \cite{Takahashi1977} gives
\begin{equation}
e(h_x)=e_0+a_2 h_x^2+a_4 h_x^4+a_6 h_x^6+a_8 h_x^8+O(h_x^{10})\ ,
\end{equation}
with $e_0=-|\epsilon|-J$ and
\begin{align}
a_2 &=-\frac{1}{2A}\\
a_4 &=\frac{|\epsilon|-J}{8BA^3}\\
a_6 &=-\frac{4J^3-11J^2|\epsilon|-10J|\epsilon|^2+3|\epsilon|^3}{16B^2A^5C}\\
a_8 &=\frac{-100J^6+460J^5|\epsilon|+1607J^4|\epsilon|^2+591J^3|\epsilon|^3-1157J^2|\epsilon|^4-609J|\epsilon|^5+90|\epsilon|^6}{128B^3DA^7C^2}\ .
\end{align}
\noindent The induced uniform magnetization is given by
\begin{align}
m_x(h_x)=\,&\frac{h_x}{A}-\frac{|\epsilon|-J}{2BA^3}h_x^3+\frac{3(3|\epsilon|^3-10|\epsilon|^2J-11|\epsilon|J^2+4J^3)}{8B^2A^5C}h_x^5\notag\\
&-\frac{90|\epsilon|^6-609|\epsilon|^5J-1157|\epsilon|^4J^2+591|\epsilon|^3J^3+1607|\epsilon|^2J^4+460|\epsilon|J^5-100J^6}{16B^3DA^7C^2}h_x^7+O(h_x^9)\ .
\label{eq:app_mh_expansion}
\end{align}
Inverting Eq.~\eqref{eq:app_mh_expansion} gives
\begin{align}
h_x(m_x)=\,&A\,m_x+\frac{A(|\epsilon|-J)}{2B}m_x^3+\frac{3|\epsilon|A(3|\epsilon|^2+2|\epsilon|J+9J^2)}{8B^2C}m_x^5\notag\\
&+\frac{A(90|\epsilon|^6+255|\epsilon|^5J+427|\epsilon|^4J^2-177|\epsilon|^3J^3+71|\epsilon|^2J^4+220|\epsilon|J^5-4J^6)}{16B^3DC^2}m_x^7+O(m_x^9)\ .
\label{eq:app_hm_expansion}
\end{align}
The Landau free energy of the short-range chain is the Legendre transform
\begin{equation}
f_0(m_x)=e(h_x)+h_xm_x\ ,
\end{equation}
where $h_x$ is eliminated in favor of $m_x$ through Eq.~\eqref{eq:app_hm_expansion}. Up to eighth order, one finds
\begin{equation}
f_0(m_x)=e_0+\frac{A}{2}m_x^2+v_4m_x^4+v_6m_x^6+v_8m_x^8+O(m_x^{10})\ ,
\end{equation}
with
\begin{align}
v_4 &=\frac{A(|\epsilon|-J)}{8B}\\
v_6 &=\frac{|\epsilon|A(3|\epsilon|^2+2|\epsilon|J+9J^2)}{16B^2C}\\
v_8 &=\frac{A(90|\epsilon|^6+255|\epsilon|^5J+427|\epsilon|^4J^2-177|\epsilon|^3J^3+71|\epsilon|^2J^4+220|\epsilon|J^5-4J^6)}{128B^3DC^2}\ .
\end{align}
The collective interaction contributes only to the quadratic term, such that
\begin{equation}
f_{\mathrm{eff}}(m_x)=f_0(m_x)+\lambda m_x^2=e_0+r\,m_x^2+v_4m_x^4+v_6m_x^6+v_8m_x^8+O(m_x^{10})\ ,
\end{equation}
where
\begin{equation}
r=\lambda+\frac{A}{2}=\lambda+\frac{|\epsilon|+2J}{2}\ .
\end{equation}
For the Dicke-Ising chain, $\lambda=-g^2/\omega$, so that
\begin{equation}
r=-\frac{g^2}{\omega}+\frac{|\epsilon|+2J}{2}\ .
\end{equation}

\end{document}